\documentclass[twocolumn,prb,amsmath,superscriptaddress]{revtex4}
\usepackage{graphicx}

\begin{document}

\title{Magnetic bubbles in FePd thin films near saturation}

\author{Thomas Jourdan}
\affiliation{CEA, INAC, SP2M, F-38054 Grenoble, France}
\affiliation{CEA, DEN, Service de Recherches de M\'etallurgie Physique, F-91191 Gif-sur-Yvette, France}
\author{Aur\'elien Masseboeuf}
\affiliation{CEA, INAC, SP2M, F-38054 Grenoble, France}
\affiliation{CNRS/CEMES, 29 rue Jeanne Marvig, 31055 Toulouse Cedex, FR}
\author{Fr\'ed\'eric Lan\c{c}on}
\affiliation{CEA, INAC, SP2M, F-38054 Grenoble, France}
\author{Pascale Bayle-Guillemaud}
\affiliation{CEA, INAC, SP2M, F-38054 Grenoble, France}
\author{Alain Marty}
\affiliation{CEA, INAC, SP2M, F-38054 Grenoble, France}

\date{\today}

\begin{abstract}
  The structure of domain walls delimiting magnetic bubbles in L1$_0$
  FePd thin layers is described on the basis of Lorentz transmission
  electron microscopy (LTEM) and multiscale magnetic
  simulations. Images obtained by high resolution LTEM show the
  existence of magnetization reversal areas inside domain walls,
  called vertical Bloch lines (VBL). Combining these observations and
  multiscale simulations on various geometries, we can identify the
  structure of these VBL, notably the presence or not of magnetic singularities.
\end{abstract}

\maketitle

\section{Introduction}
\label{sec:introduction}

FePd alloys with L1$_0$ structure deposited in thin layers have
attracted much attention because of their very high perpendicular
anisotropy, which is a key property for magneto-optical recording and
for high density magnetic storage. Recently, alloys with a
perpendicular anisotropy have been used in spin-valves, where they are
used as the polarizer and as the free layer that should be
reversed\cite{seki06}. It has been shown that in such
devices\cite{seki08} or in magnetic tunnel junctions,
the reversal of the free layer occurs through the nucleation of a
reversed domain followed by the propagation of a domain wall.


Near the saturation, the band domain structure in FePd layers
transforms into a lattice of magnetic bubbles\cite{hubert98}, which
remains stable at high fields. In some bubbles the Bloch-like walls have
different polarities, separated by segments called vertical Bloch
lines (VBL). In the present work we analyze the role of VBL on the
shape of these magnetic bubbles.

VBL were much studied in the 1980s in garnets, both experimentally and
numerically. Typical parameters for these garnets are $K =
10^3~\textrm{J.m}^{-3}$, $M_s = 1.4\times 10^4~\textrm{A.m}^{-1}$ and
$A = 1.3\times 10^{-12}~\textrm{J.m}^{-1} $, so that the domain wall
width is $\delta = \pi\sqrt{A/K} \approx 0.1~\mu\textrm{m}$. This
large value, compared to the domain wall width in FePd of around 8~nm,
makes possible the optical observation of domain walls and
VBL\cite{thiaville90}. In FePd, a higher resolution is necessary to
probe the sample, which can be reached by Lorentz transmission electron
microscopy (LTEM). Extensive analytical and numerical studies have also
been performed in VBL in
garnets\cite{slonczewski74,hubert74,nakatani88,miltat89}. Given the
high value of the quality factor $Q = 2K/(\mu_0 M_s^2) \approx 8$, a
common assumption in the models is $Q \gg 1$, which notably permits to
use a local approximation of the demagnetizing field and thus
simplifies the calculations. This assumption is \emph{a priori} not
valid in the case of FePd, which exhibits smaller values of $Q$ of the
order of 1.6.

In the present work we performed high resolution imaging of domain
walls in magnetic bubbles in FePd thin layers, using Lorentz
microscopy, to highlight their magnetic configuration. In particular
we describe the influence of VBL on the shape of the bubbles. We also
show the results of multiscale simulations that provide an explanation
for these observed shapes.



\section{Observation of magnetic bubbles in FePd thin film}
\label{sec:observation-magnetic-bubbles-FePd}

Lorentz microscopy is now a well established method that enables
magnetic imaging with a resolution better than ten nanometers. The
simplest mode of LTEM is the observation of the overlapping of
electrons experiencing different Lorentz forces in magnetic
domains. The contrasts obtained by simply defocalizing the lens used
for imaging are called Fresnel contrasts \cite{Chapman1984}. In a
classical in-plane magnetization configuration, Fresnel contrasts
appear on the domain walls position due to the overlapping of
electrons coming from two opposite domains. In the particular case of
FePd, where magnetization is mainly out-of-plane, the contrasts can be
obtained by tilting the sample \cite{Aitchison2001}. This enables the
magnetization inside the domain to act on the electron beam and to
produce traditional Fresnel contrasts located on the domain
walls. Otherwise contrasts can be produced by the domain walls
themselves if the layer is thick enough and if the amount of in-plane
magnetization in the wall is large enough\cite{Masseboeuf2009}
(\textit{i.e.} to reach the LTEM sensitivity of about 10~nm.T). This
was the case for our samples, so we have performed Fresnel
observations of Bloch walls without tilting the FePd layers. The
microscope used in these observations was a JEOL 3010 fitted in with a
Gatan imaging filter for contrast enhancement\cite{Dooley1997}. The
images displayed in this letter have been also filtered by a Fourier
approach to enhance the contrasts localized on the domain walls. The
magnetization was performed using the objective lens, calibrated with
a Hall Probe. The sample was prepared by Molecular Beam Epitaxy on MgO
[001] substrate. The magnetic stacking is decomposed in two layers: a
``soft'' layer of 17~nm FePd$_2$ having a vanishing anisotropy is
deposited before a 37~nm-L1$_0$ layer of FePd. Details can be found
in Ref.~\onlinecite{Masseboeuf2008}. The sample was prepared for TEM
observation with a classical approach: mechanical polishing and ion
milling.

\begin{figure*}[htbp]
  \centering
  \includegraphics[width=\linewidth]{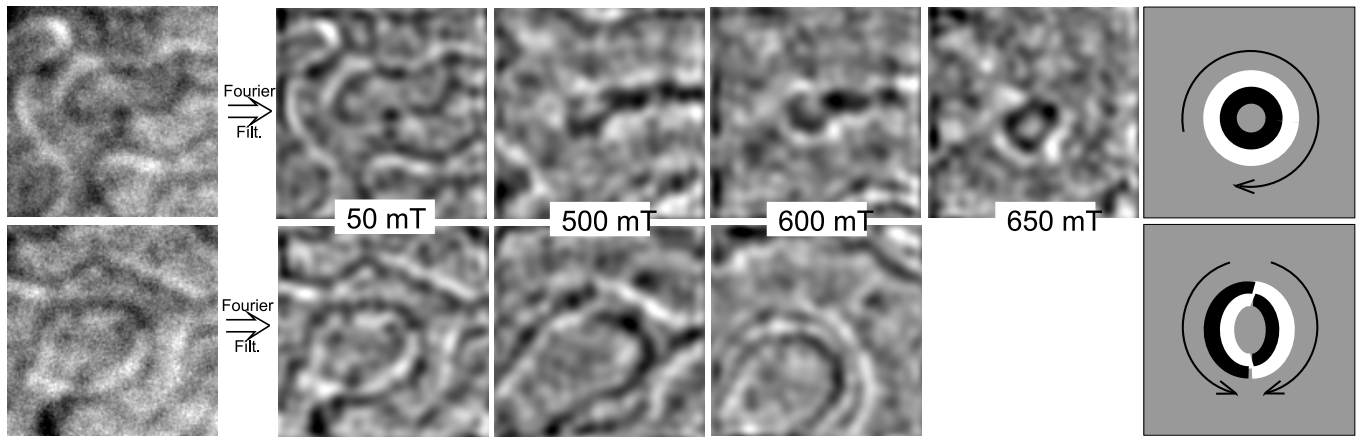}  
  \caption{Magnetization process on FePd thin film. The two rows
    present two different areas in the film. Both of them present a
    magnetic bubble state just before saturation. Left images are raw
    datas while the other images are enhanced by Fourier
    filtering. Right images are simple schemes to highlight the contrast observed in the last step of magnetization process. Arrows point out the direction of magnetic induction in bubbles. Images are 500 $\times$ 500~nm.}
  \label{LTEM1}
\end{figure*}

Fig.~\ref{LTEM1} shows two different areas of the foil during the
magnetization process. We observe couples of black and white contrasts
corresponding to the Bloch walls\cite{Masseboeuf2009}. These pictures
have been obtained for increasing applied fields. We should notice
that upon 500~mT the quality of the images decreases due to the action
of the objective lens on the image formation. Nevertheless it is
possible to follow the shape of the domains during the magnetization
process (enhanced here by Fourier filtering). We observe in both cases
that a magnetic domain collapses to a bubble state. Attention can thus
be paid on the chirality of the Bloch wall. The chirality (sense of
the magnetization inside the Bloch wall) is directly linked to the
Fresnel contrast: the wall chirality of a black/white contrast and the
chirality of a white/black contrast are opposite. Knowing this, the
observation of the two magnetic bubbles presented in the right images
of Fig.~\ref{LTEM1} gives some information on the magnetization inside
the domain walls of the bubbles. The first bubble presents a
continuous domain wall, swirling all around the bubble, whereas the
other one exhibits two different parts with the same magnetization
orientation. In the latter configuration, the magnetization inside the
domain wall experiences two rotations of 180$^{\circ}$ localized
at the top and the bottom of the bubble. These switching areas are
known as vertical Bloch lines (VBL). One can notice the main
difference in the two bubble shapes: the first one is almost round
while the second bubble seems to be slightly elongated along the
vertical direction.

To confirm the role of VBL on the bubble shape, we have thus simulated
the inner structure of domain walls containing VBL.

\section{Simulation of domain walls with vertical Bloch lines}
\label{sec:simulation-VBL}

The numerical simulation of magnetic bubbles is not a tractable
problem with standard codes. Indeed it requires to handle
large systems whose size is related to the size of the bubbles, but
with regions where the magnetization varies rapidly in space, such as
domain walls and all their substructures. Considering all regions with
the same level of refinement is clearly not well adapted to such a
multiscale problem and leads to a high computational effort. The same
level of accuracy can be reached with a coarser mesh in uniformly
magnetized regions.

In this work we used a multiscale code (Mi\_$\mu$Magnet) based on an adaptive mesh
refinement technique, as well as on a mixed atomistic-micromagnetic
approach, to achieve both precision and computational
efficiency\cite{jourdan08}. Given the large size of the systems we
envisage here, the code was only used in its micromagnetic mode. It
has been recently shown that micromagnetic calculations can be applied
to singularities appearing in VBL, called Bloch Points
(BP)\cite{thiaville03}. In all calculations the mesh step is kept
lower than half the exchange length.

Parameters are chosen in agreement with experimental
measurements\cite{gehanno97-2}: the saturation magnetization,
anisotropy constant and exchange stiffness are $M_s =
10^6~\textrm{A.m}^{-1}$, $K=10^6~\textrm{J.m}^{-3}$, and $A = 7\times
10^{-12}~\textrm{J.m}^{-1}$. With such parameters, the exchange length
is $l_{ex} = \sqrt{2A/(\mu_0M_s^2)}=3.3~\textrm{nm}$.

Two types of computations have been carried out. First we investigate
the properties of a straight domain wall containing a VBL. Secondly we
study the role of VBL on the shape of the magnetic bubbles in FePd
layers.

\subsection{Vertical Bloch lines in straight domain walls}
\label{sec:VBL-straight-DW}

\newcommand{\mup}[4]{%
\filldraw[black, very thick] (#1,#2) circle (#3);
\draw[black, very thick] (#1,#2) circle (#4);
}

\newcommand{\mdn}[3]{%
\draw[black, very thick] (#1,#2) circle (#3);
\draw[black, very thick] (#1,#2) -- +(45:#3);
\draw[black, very thick] (#1,#2) -- +(135:#3);
\draw[black, very thick] (#1,#2) -- +(225:#3);
\draw[black, very thick] (#1,#2) -- +(315:#3);
}

\newcommand{\cpl}[3]{%
\draw[black, very thick] (#1,#2) -- +(#3,0);
\draw[black, very thick] (#1,#2) -- +(-#3,0);
\draw[black, very thick] (#1,#2) -- +(0,#3);
\draw[black, very thick] (#1,#2) -- +(0,-#3);
}

\newcommand{\cms}[3]{%
\draw[black, very thick] (#1,#2) -- +(#3,0);
\draw[black, very thick] (#1,#2) -- +(-#3,0);
}

\begin{figure}[htbp]
  \centering
  \includegraphics[width=0.8\linewidth]{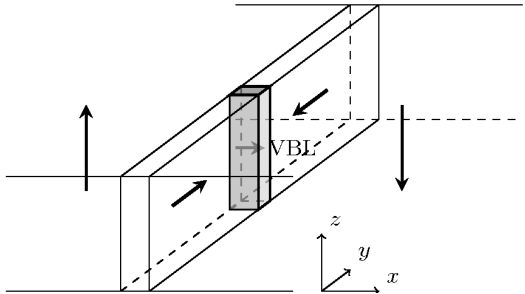}  
  \caption{Schematic representation of the system used to study the
    structure of VBL in domain walls.}
  \label{fig-simulation-VBL-straight-DW}
\end{figure}
\begin{figure}[htbp]
  \centering
  \includegraphics[width=\linewidth]{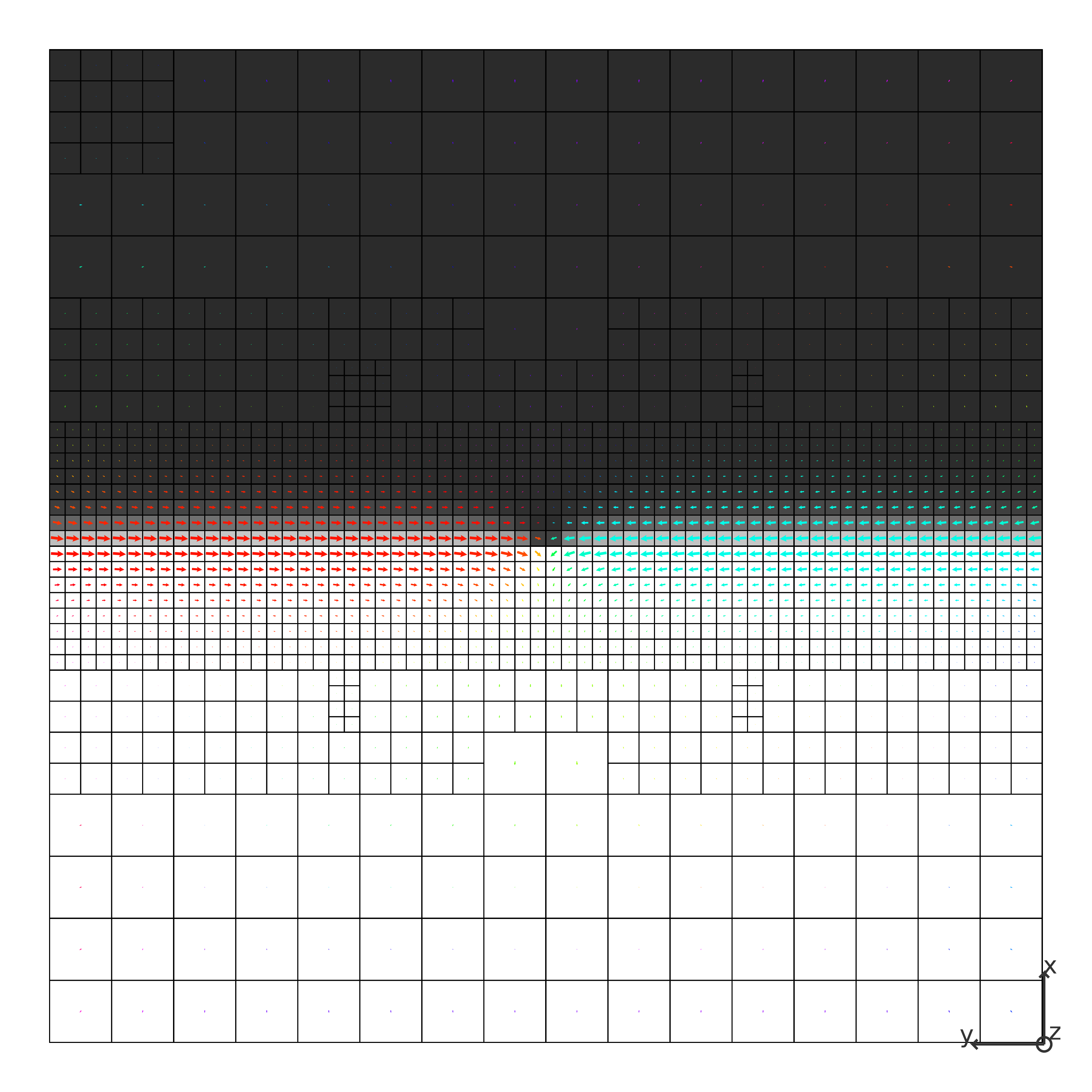}
  \caption{(Color online) Cross-section of the whole system obtained
    from a multiscale simulation (Bloch wall, with a vertical Bloch
    line containing a Bloch point). The orientation of the
    magnetization for the in-plane component is given by the arrows
    and the color wheel, and by the grayscale for the out of plane
    contribution (along $z$). The norm of the arrows is proportional
    to the in plane magnetization. The lateral size of the system is
    $110~\textrm{nm}\times 110~\textrm{nm}$.}
  \label{fig-vbl-BP-all}
\end{figure}

The system used here contains a domain wall with a
single VBL (Fig.~\ref{fig-simulation-VBL-straight-DW}). The lateral
size of the system is $110~\textrm{nm}\times 110~\textrm{nm}$ and the
thickness of the layer is varied between 11.3 to 37.6~nm. No periodic
boundary conditions are used, because this would involve a second VBL
along $y$ and a second domain wall along $x$. A view of such a system
is given in Fig.~\ref{fig-vbl-BP-all}.

\begin{figure}[htbp]
  \centering
  \includegraphics[width=\linewidth]{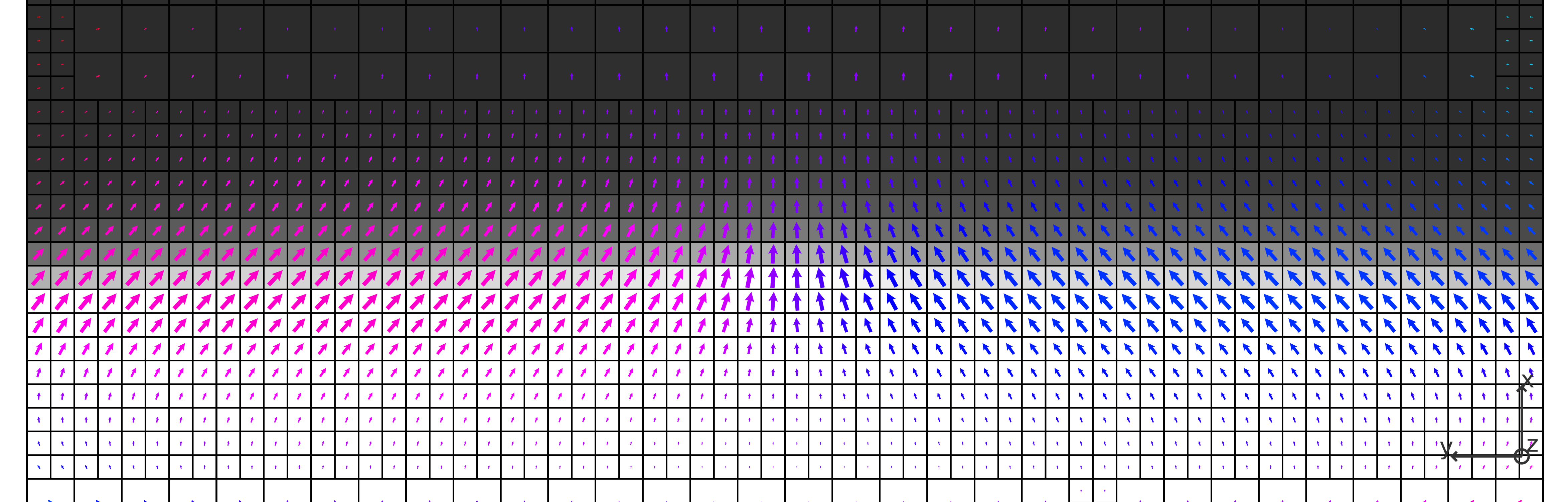}
  
  \includegraphics[width=\linewidth]{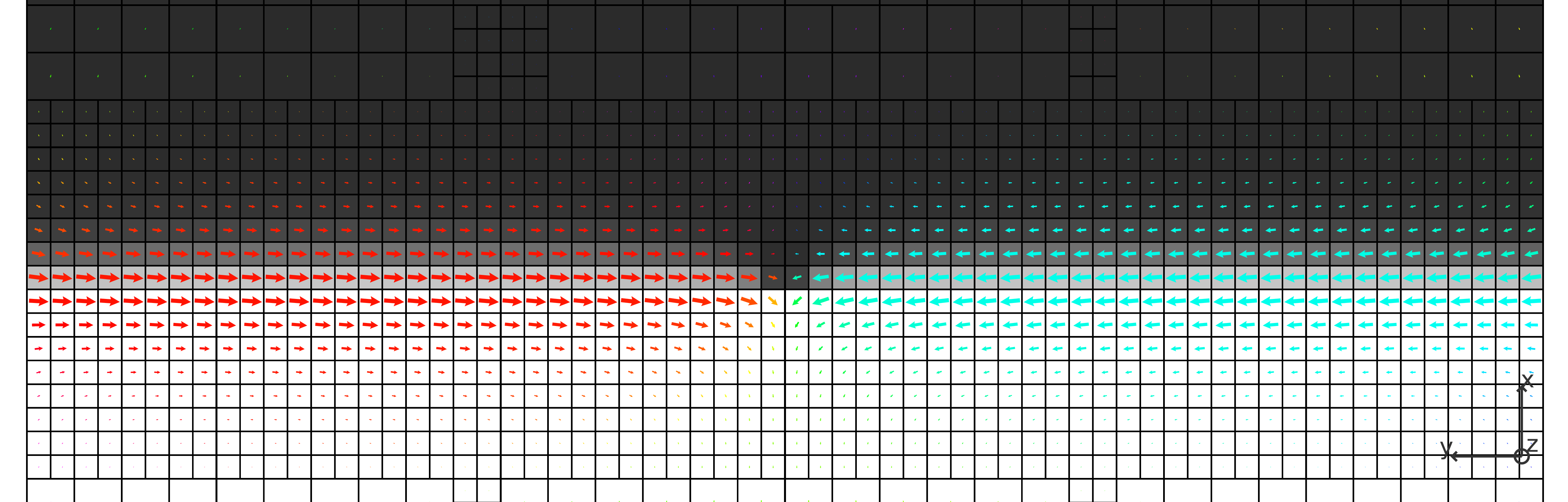}

  \includegraphics[width=\linewidth]{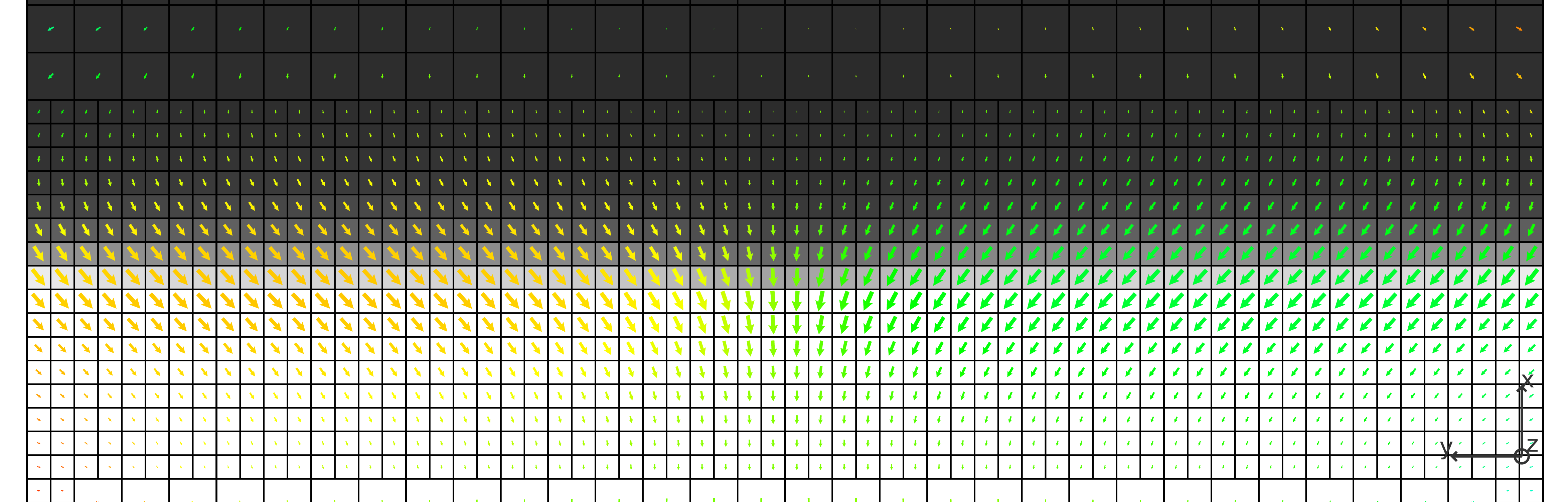}
  \caption{(Color online) Cross-section along the planes $z=h$, $z=h/2$ and $z=0$
    (from top to bottom) with a VBL containing a BP.}
  \label{fig-vbl-BP}
\end{figure}

\begin{figure}[htbp]
  \centering
  \includegraphics[width=\linewidth]{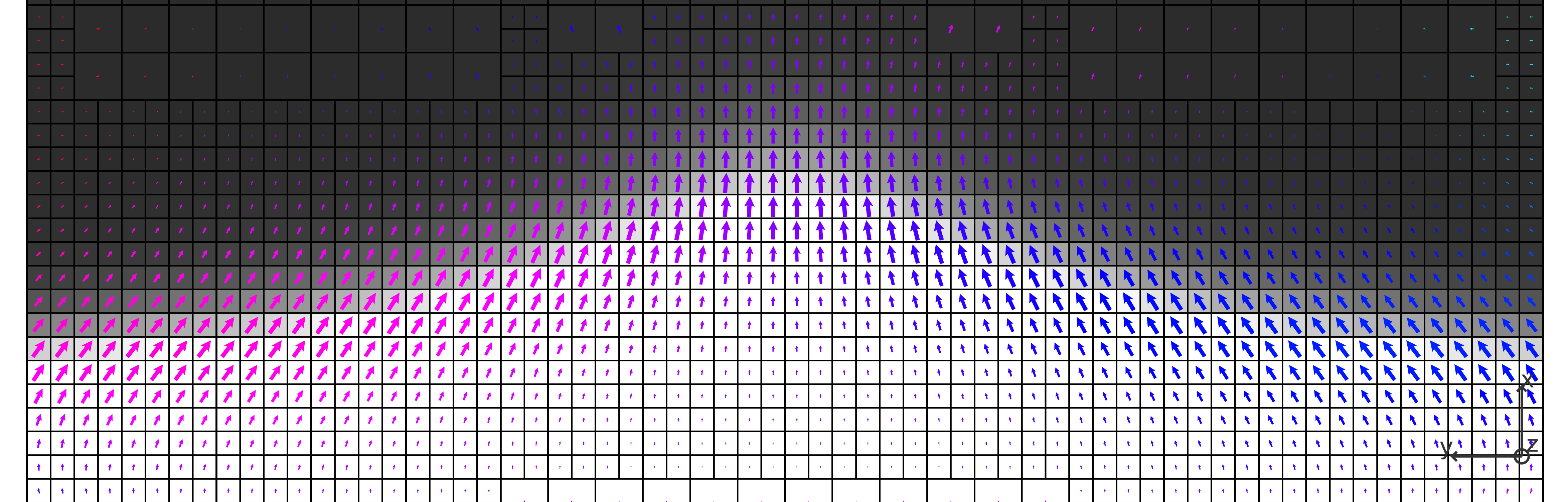}
  
  \includegraphics[width=\linewidth]{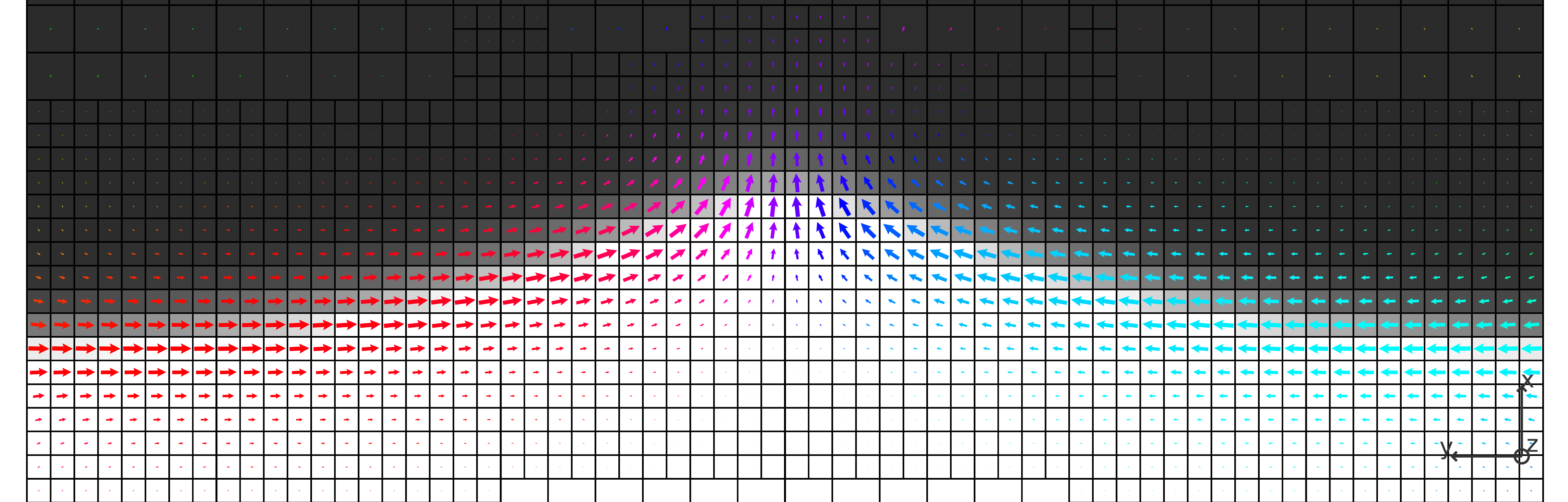}

  \includegraphics[width=\linewidth]{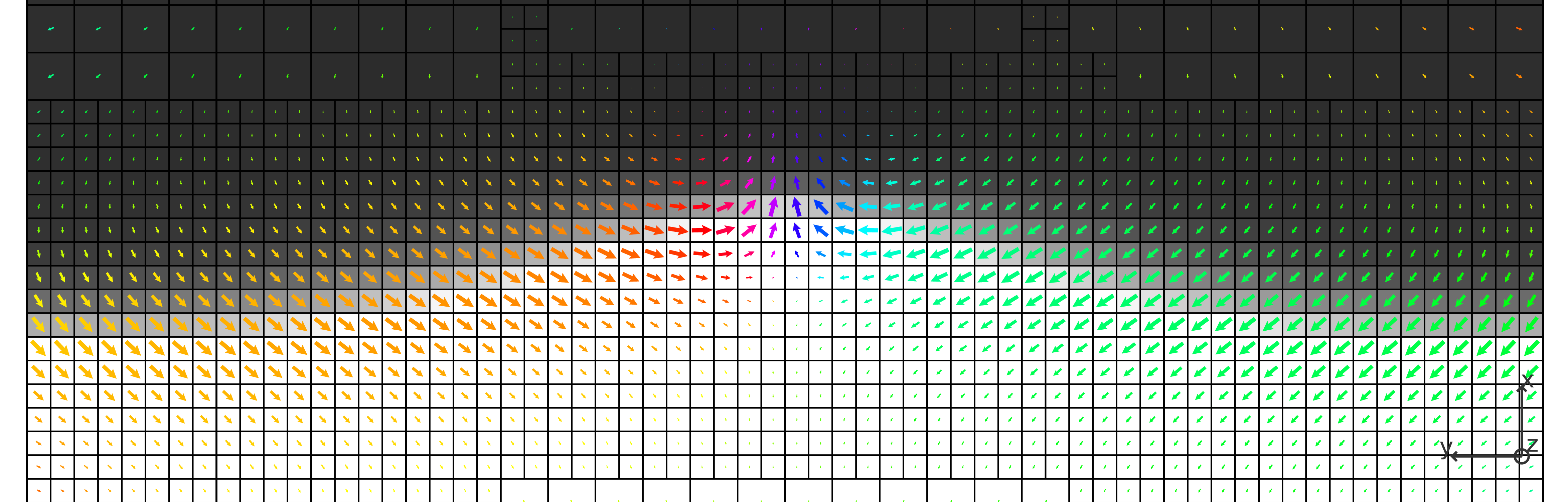}
  \caption{(Color online) Cross-section along the planes $z=h$, $z=h/2$ and $z=0$
    (from top to bottom) with a VBL containing no BP.}
  \label{fig-vbl-noBP}
\end{figure}

\begin{figure}[htbp]
  \centering
  \includegraphics[width=\linewidth]{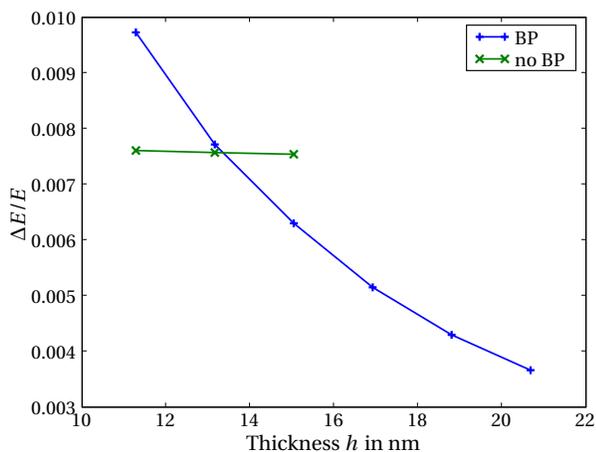}
  \caption{(Color online) Relative energy of a system with and without a BP in the
    VBL. The reference energy is given by the system containing a
    domain wall without a VBL. The decrease of the energy when a BP is
    present is mainly due to the dipolar term.}
  \label{fig-stability-BP}
\end{figure}

For all the values of the thickness $h$ that we envisage, we consider
configurations with and without a BP (Fig.~\ref{fig-vbl-BP}
and~\ref{fig-vbl-noBP}). The configuration without a BP can be
stabilized only for a thickness lower than 15~nm and becomes
energetically favorable below a critical thickness of around 13~nm
(Fig.~\ref{fig-stability-BP}). For thicknesses larger than 15~nm, well
defined N\'eel caps are present due to the dipolar field created by
the domains and a BP nucleates on the surface where the magnetization
rotates of nearly 360$^\circ$ (Fig.~\ref{fig-vbl-noBP}, at $z =
0$). It must be noted that the critical thickness is around
$4~l_{ex}$, which is significantly lower than the value $7.3~l_{ex}$
found by a variational method\cite{hubert76}. Indeed this method,
based on a local approximation of the dipolar field, is well justified
if $Q \gg 1$ but does not hold in our case.


An interesting feature of the domain wall containing a VBL without a
BP is the so-called buckling of the magnetization near the line. This
buckling was already described in garnets, where it was ascribed to
the diminution of the magnetic charges created by the variation of the
magnetization in the direction orthogonal to the wall\cite{miltat89}
(which we note $x$ here). These charges are called $\pi$ charges or
dipolar charges, in analogy with $\pi$ orbitals, because positive
charges are associated with negative charges.

Analytical models, based on the assumption $Q \gg 1$, and two
dimensional simulations, predict a much smaller value of the
buckling\cite{miltat89}. Given our material parameters, it would be less than 2~nm,
whereas it is around 10~nm for all the thicknesses we have
considered. Three dimensional simulations for $Q = 7.7$ and thick
garnet layers ($h \approx 50~l_{ex}$) also give a tiny buckling. In
this case, a tilt of the wall is observed in the $x-z$
plane that provides a compensation for the charges associated with the
variation of the magnetization along $y$ (called $\sigma$ or monopolar
charges)\cite{thiaville91}.

\begin{figure}[htb]
  \centering
    \includegraphics[width=\linewidth]{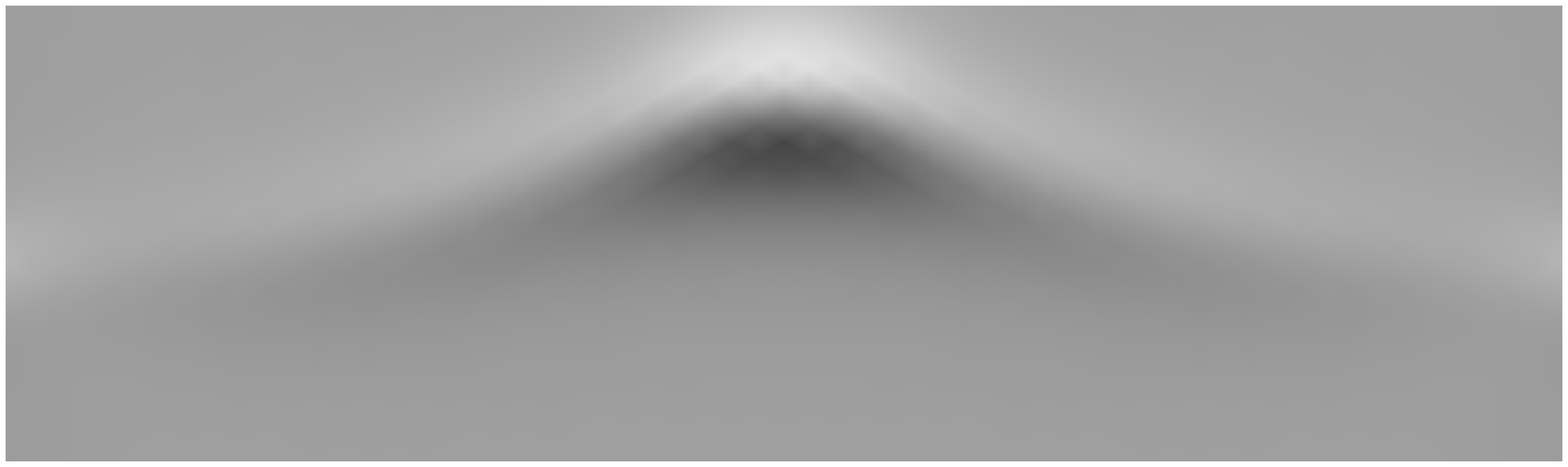}
    
    \vspace{0.5cm}
    
    \includegraphics[width=\linewidth]{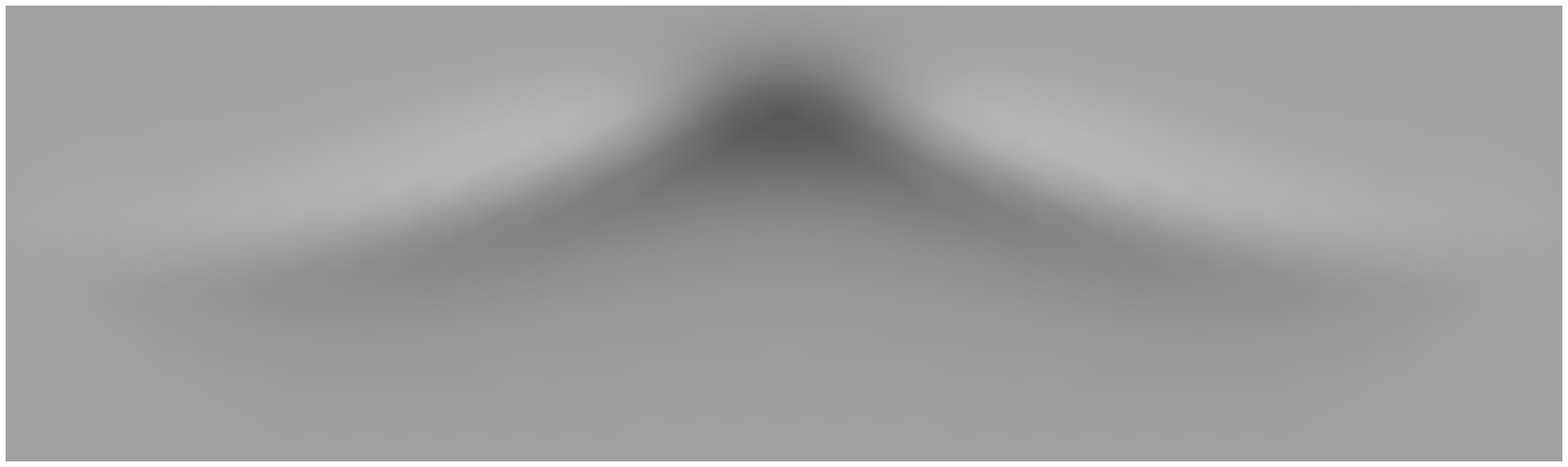}  
    
    \vspace{0.5cm}
    
    \includegraphics[width=\linewidth]{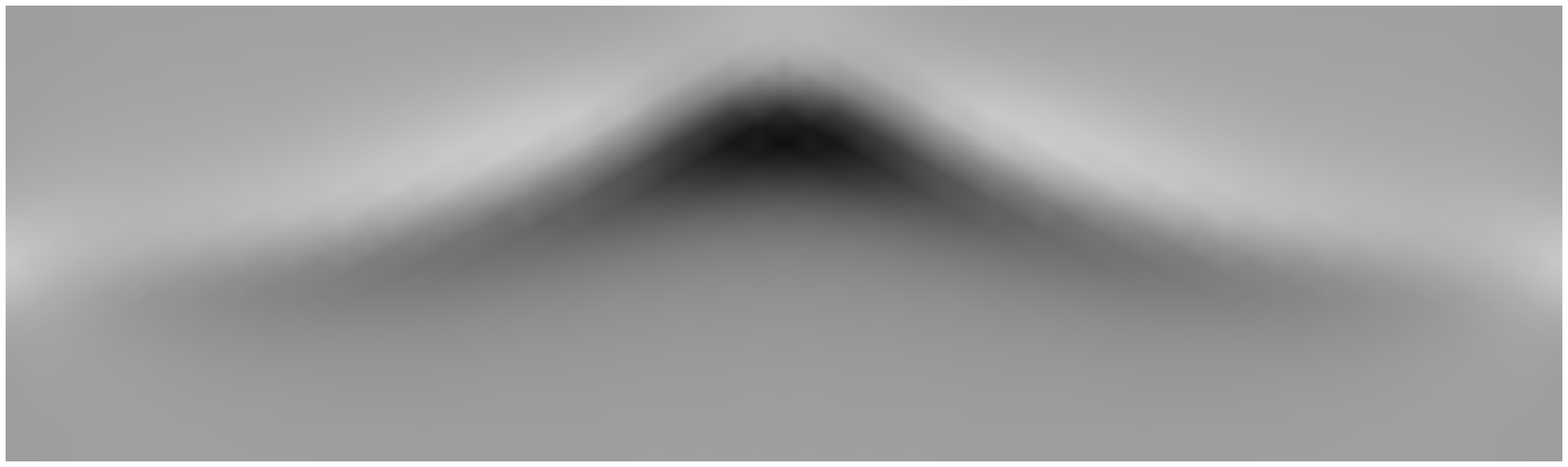}  
    
    \includegraphics[width=0.8\linewidth]{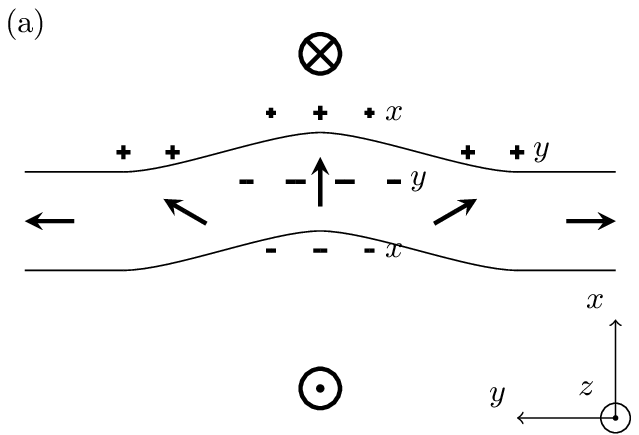}

  \caption{Magnetic charges in the plane $z=h/2$ corresponding to the
    configuration of Fig.~\ref{fig-vbl-noBP} when the transformation
    $\phi\rightarrow -\phi$ is performed on the magnetization and the
    configuration is left unrelaxed. From top to bottom: charges
    associated with the variation of $m_x$, $m_y$ and total
    charge. The charges due to the variation along $z$ are the same
    before and after the transformation and are not
    represented. Positive and negative charges are represented
    repectively by light and dark gray tones. On the schematic the
    letters $x$ and $y$ refer to charges due to the variation of $m_x$
    and $m_y$.}
  \label{fig-charges-ini}
\end{figure}

\begin{figure}[htb]
  \centering
  \includegraphics[width=\linewidth]{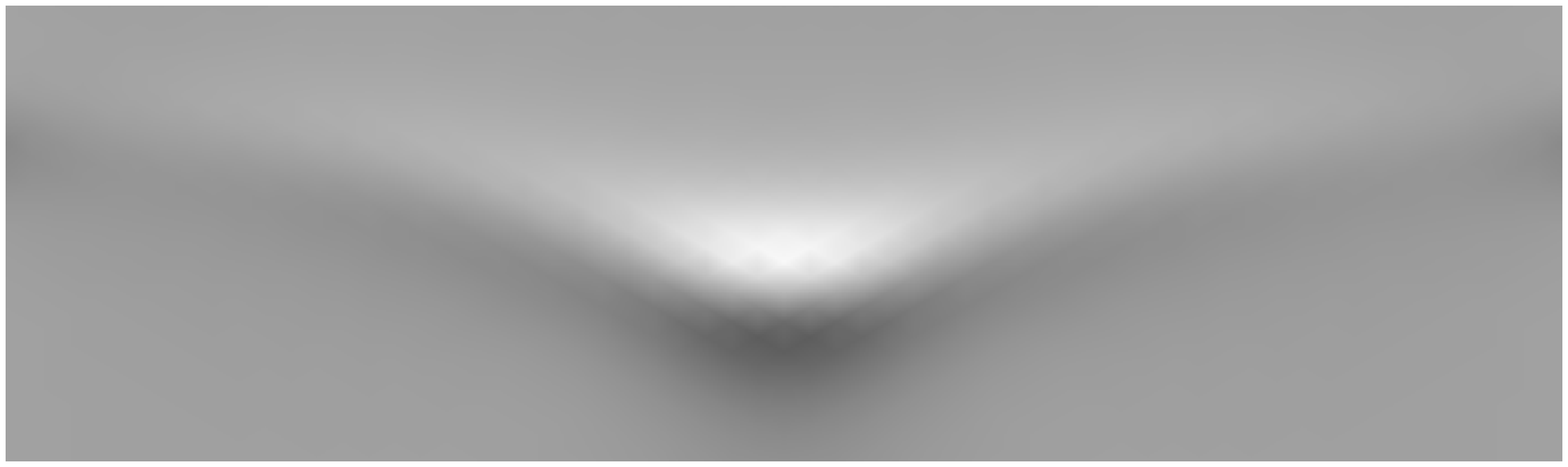}
    
  \vspace{0.5cm}
    
  \includegraphics[width=\linewidth]{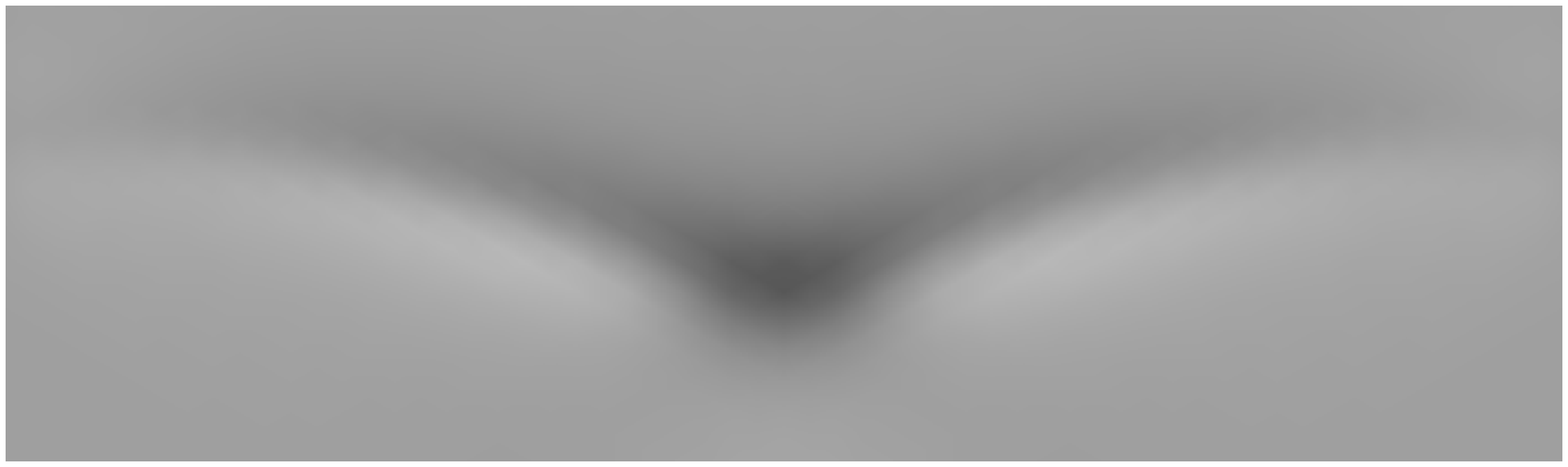}  
    
  \vspace{0.5cm}
  
  \includegraphics[width=\linewidth]{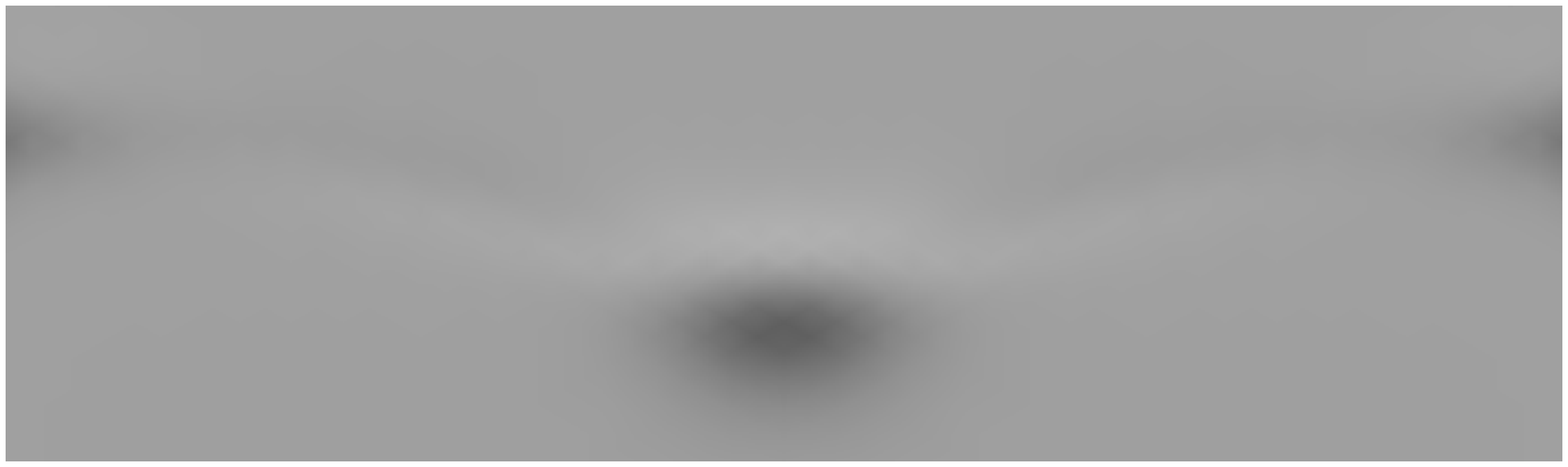}     

  \includegraphics[width=0.8\linewidth]{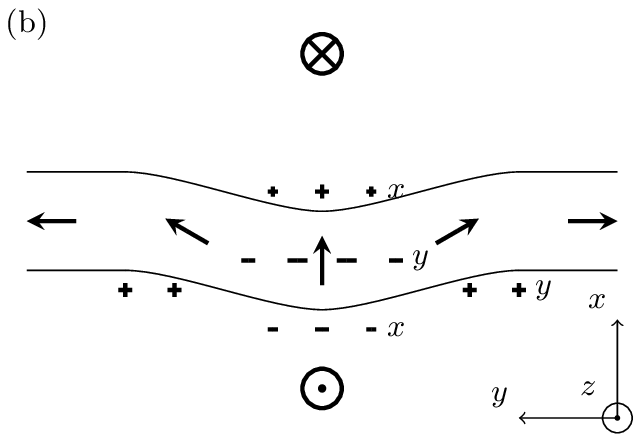}

  \caption{Magnetic charges in the plane $z=h/2$ after the relaxation
    of the configuration in Fig.~\ref{fig-charges-ini}.}
  \label{fig-charges-fin}
\end{figure}

Such a deformation is not present in our simulations. As shown in
Fig.~\ref{fig-charges-ini} and~\ref{fig-charges-fin}, the compensation
of the $\sigma$ charges is achieved by the buckling itself. It can be
noted that this buckling is due to the dipolar term, although a small
decrease of the exchange energy is also observed in the presence of
buckling. Indeed, we have represented in Fig.~\ref{fig-charges-ini} the
magnetic charges $-\partial m_x/\partial x$ ($\pi$ charges),
$-\partial m_y/\partial y$ ($\sigma$ charges) and the total charges
after a transformation $\phi \rightarrow -\phi$ on the configuration
of Fig.~\ref{fig-vbl-noBP}. The angle $\phi$ refers to the orientation of
the magnetization in the plane of the layer. The configuration after
the minimization of the energy is shown in
Fig.~\ref{fig-charges-fin}. The deformation of the domain wall has
reversed, whereas the exchange energy was invariant under the
transformation. This indicates that the deformation must be ascribed
to the compensation of $\pi$ and $\sigma$ charges, which cannot really
be distinguished, given the moderate value of $Q$. Incidentally, the
name ``$\sigma$ charges'' is not really adapted to our case given that
positive charges are associated with negative charges along $y$.

Unfortunately, no conclusions can be made from experimental datas on
the straight wall. As seen in Fig.~\ref{LTEM1}, at low fields the
domain walls in this sample are not straight enough.
On the contrary, is is possible to simulate entire bubbles and thus to
reproduce the geometry of domain walls near saturation.

\subsection{Magnetic bubbles}
\label{sec:magnetic-bubbles}

It seems reasonable to think that the deformation observed in straight
domain walls can be responsible for the distorted shape of the magnetic
bubbles. 
However, the curvature of the magnetic bubbles is such that the domain
wall cannot be considered as a straight object. The presence of two
VBL in a bubble, that bear opposite $\sigma$ charges and thus attract
themselves, may also affect the distortion. Therefore it is necessary
to perform the simulation of entire magnetic bubbles.

The system considered in these simulations contains a magnetic bubble
centered in a square of length 218~nm. Three thicknesses are
envisaged: 15~nm, 20.7~nm and 37.6~nm. Periodic boundary conditions
are used along $x$ and $y$ to simulate an array of bubbles. The
distance between the bubble's centers is thus 218~nm and is close to
the experimental value of about 250~nm. The use of an adaptive mesh
refinement technique permits to decrease the number of variables by a
factor of around 8.

Stability of bubbles is achieved for applied fields between two
critical values: if the field is too high, the bubble collapses, and
if the field is too low, the bubble transforms into a stripe domain
pattern\cite{thiele70}. For a thickness of 37.6~nm, we find that the
collapse field is between 0.6 and 0.7~T, close to the experimental
value of 0.8~T.

\begin{figure}[htbp]
  \centering
  \includegraphics[width=0.47\linewidth]{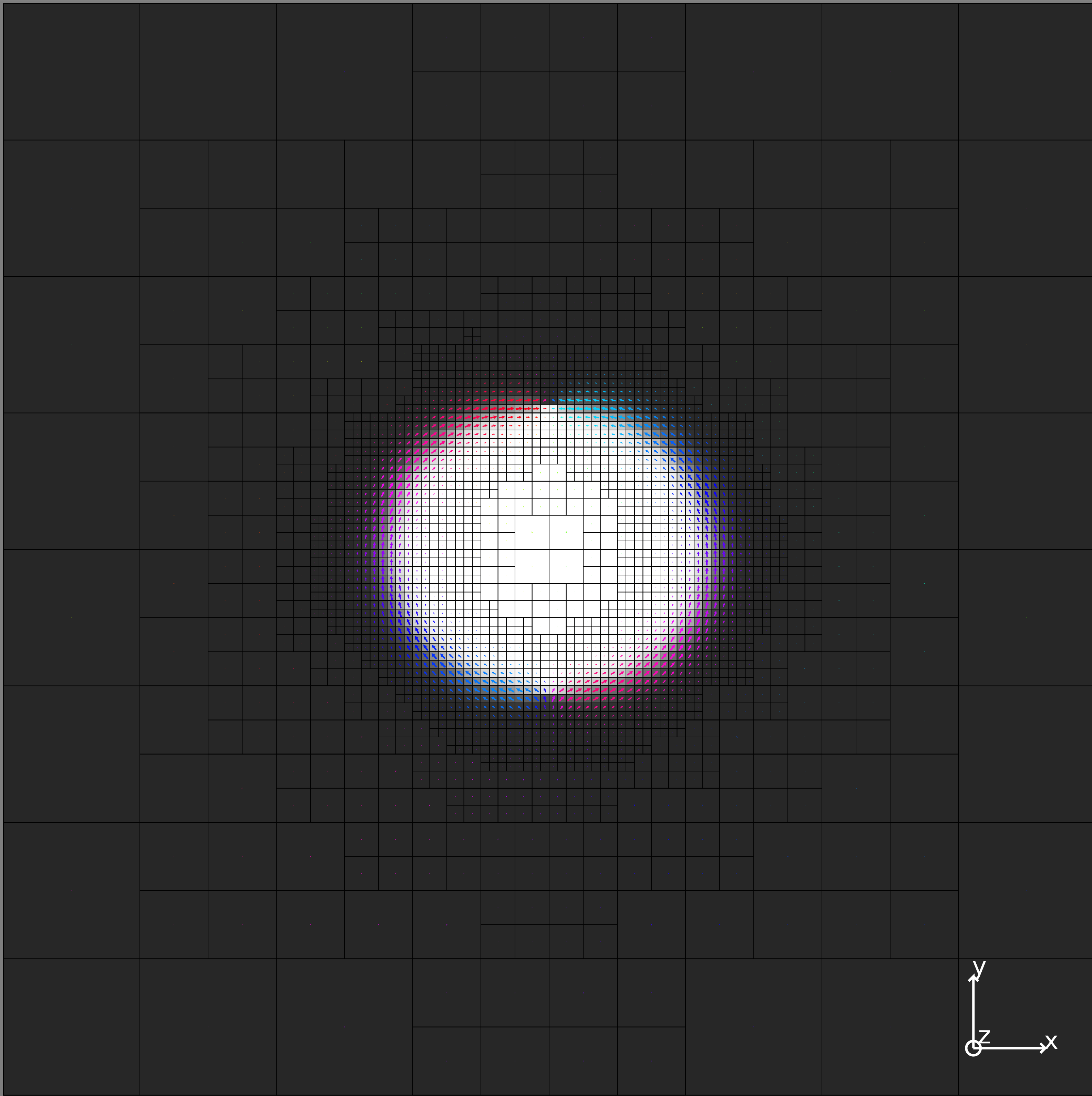}
  \hspace{\fill}
  \includegraphics[width=0.48\linewidth]{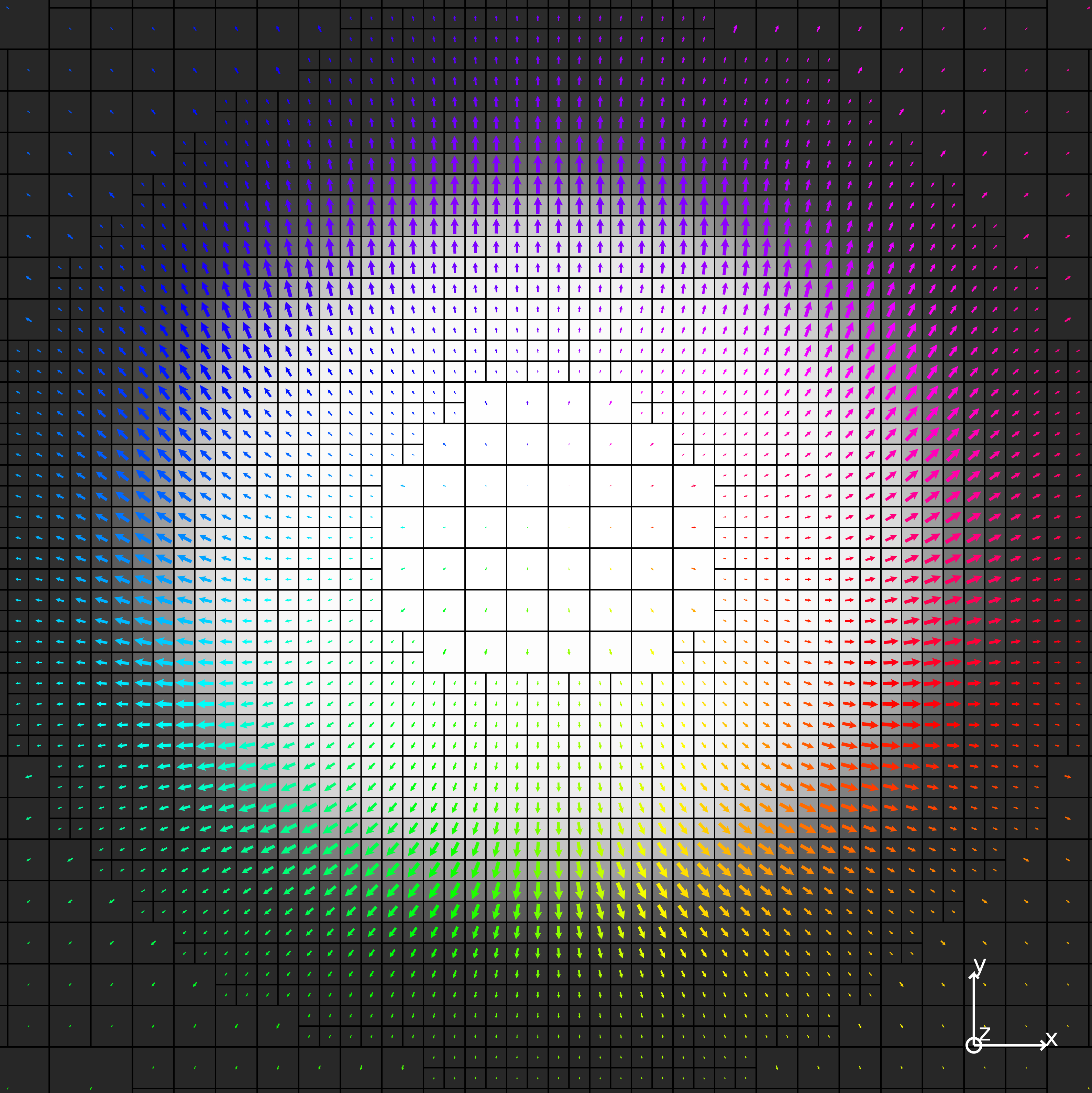}
  
  \vspace{0.2cm}
  
  \includegraphics[width=0.48\linewidth]{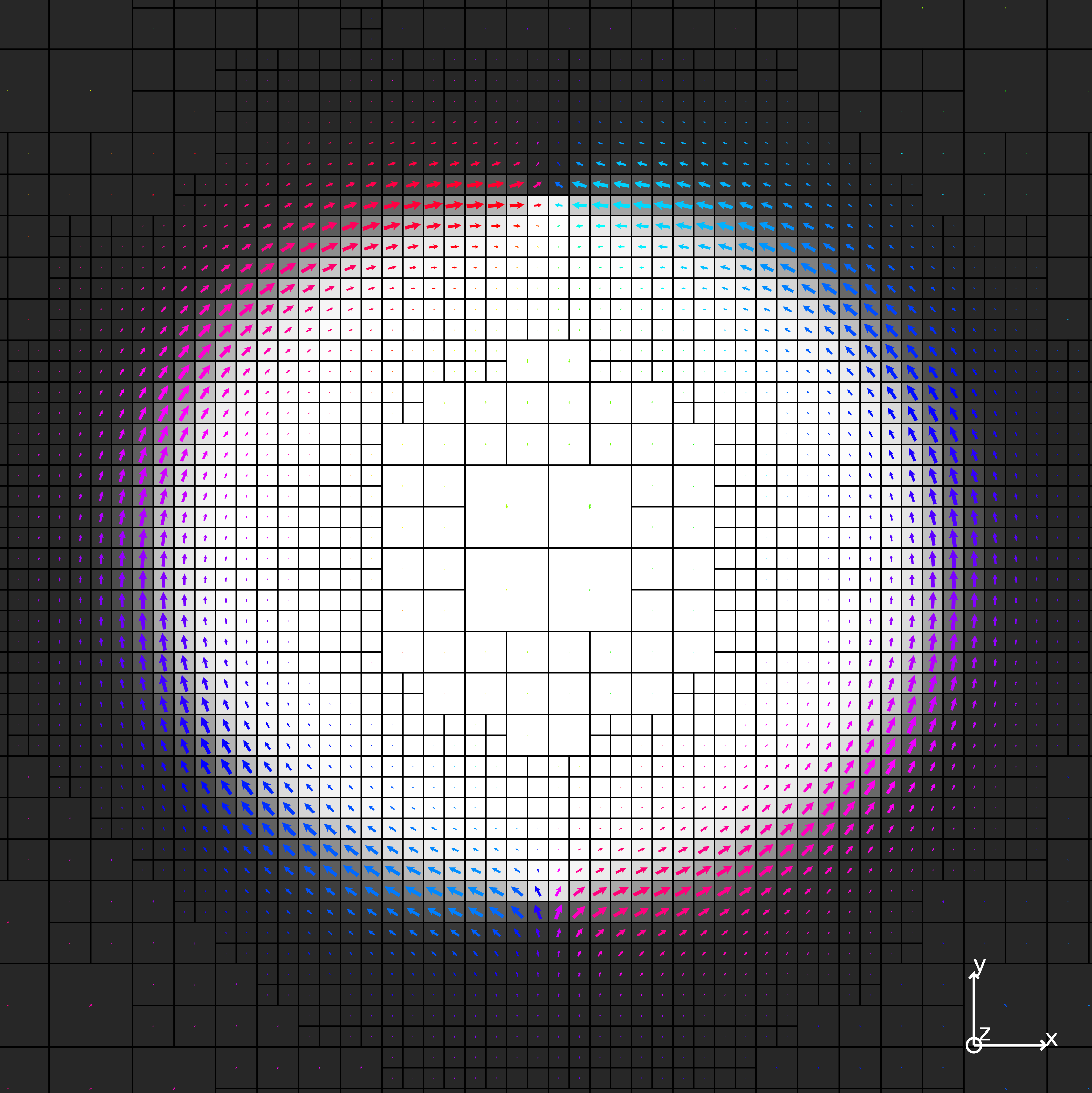}
  \hspace{\fill}
  \includegraphics[width=0.48\linewidth]{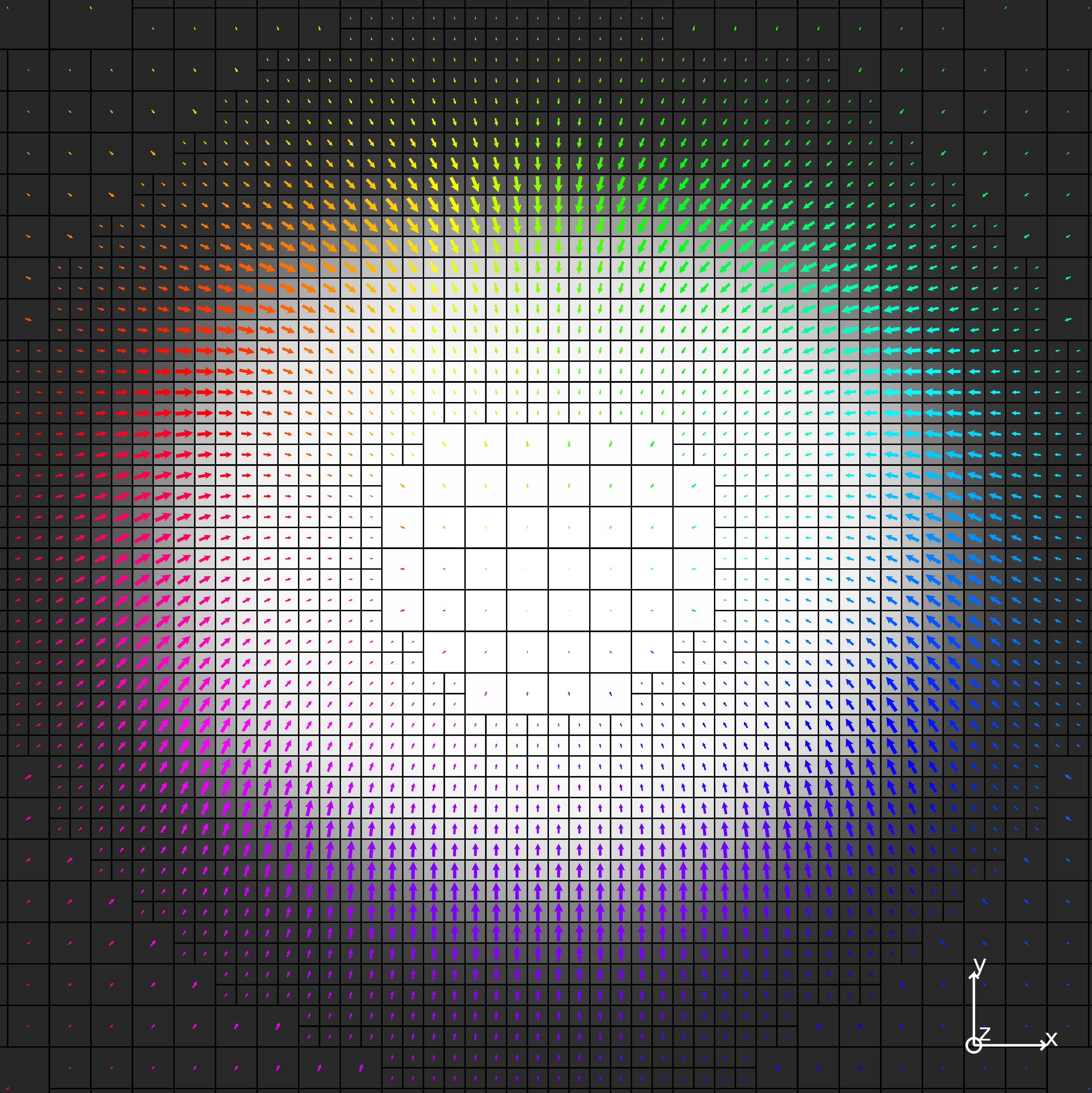}  

  \caption{(Color online) Cross-sections of a system containing two
    VBL with a BP. From top left to bottom right: whole system at
    $z=h/2$ (lateral size $218~\textrm{nm}\times 218~\textrm{nm}$),
    zoom at $z=h$, $z=h/2$ and $z=0$ (lateral size
    $90~\textrm{nm}\times 90~\textrm{nm}$). The system is 20.7~nm
    thick and a field of 0.3~T is applied. The largest cell lateral
    size is 27.3~nm, while the smallest is 1.7~nm.}
  \label{fig-bubble-with-BP}
\end{figure}

For thicknesses of 20.7~nm and 37.6~nm, it is not possible to
stabilize the configuration with two VBL without a BP. As
observed for straight domain walls, two BP nucleate because
of the dipolar field. The bubbles with VBL containing BP are
found to be almost circular (Fig.~\ref{fig-bubble-with-BP}). The small
distortion may be ascribed either to the interaction between the two
VBL which possess opposite charges, or to a local stiffness due to the
presence of the BP.

\begin{figure}[htbp]
  \centering
  \includegraphics[width=0.48\linewidth]{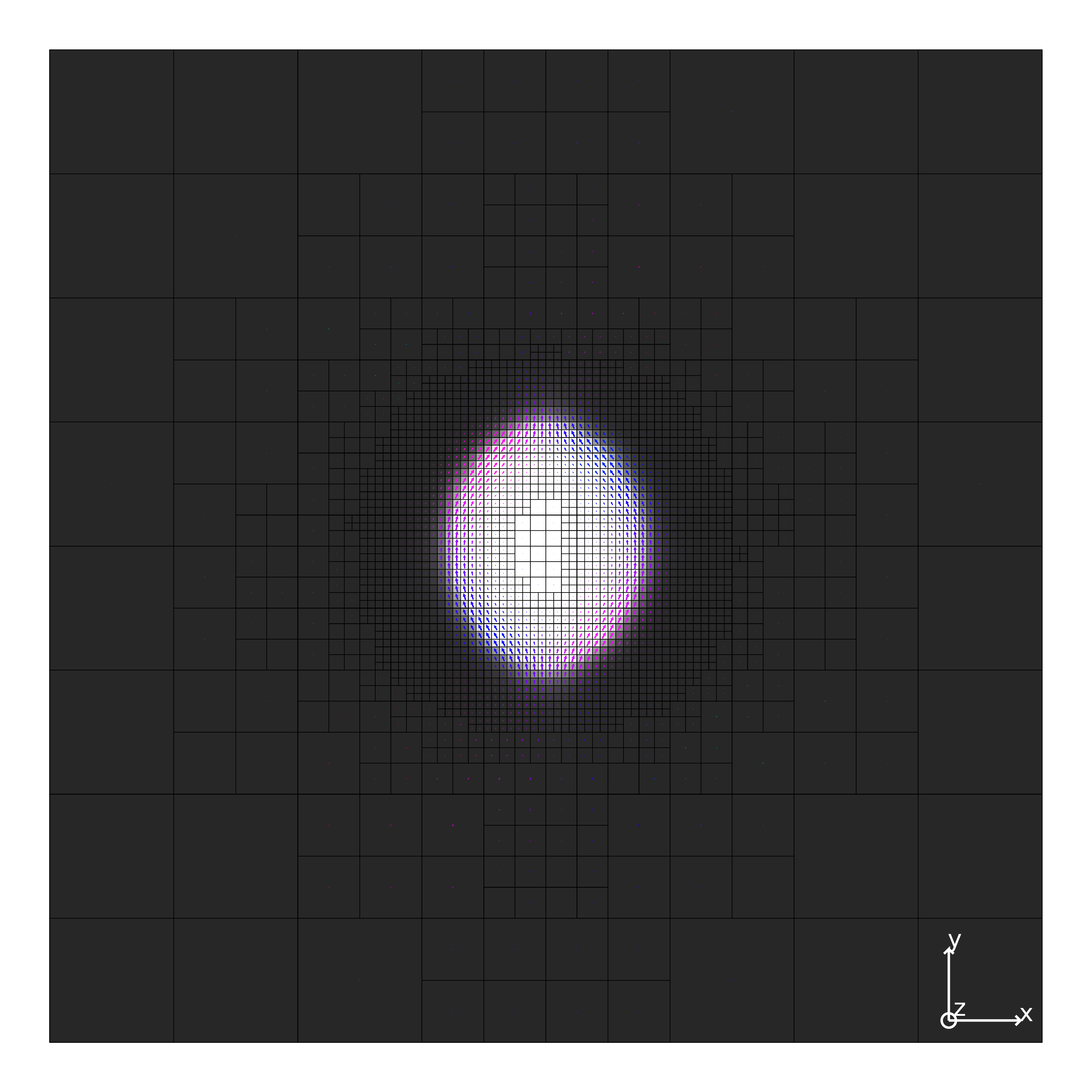}
  \hspace{\fill}
  \includegraphics[width=0.48\linewidth]{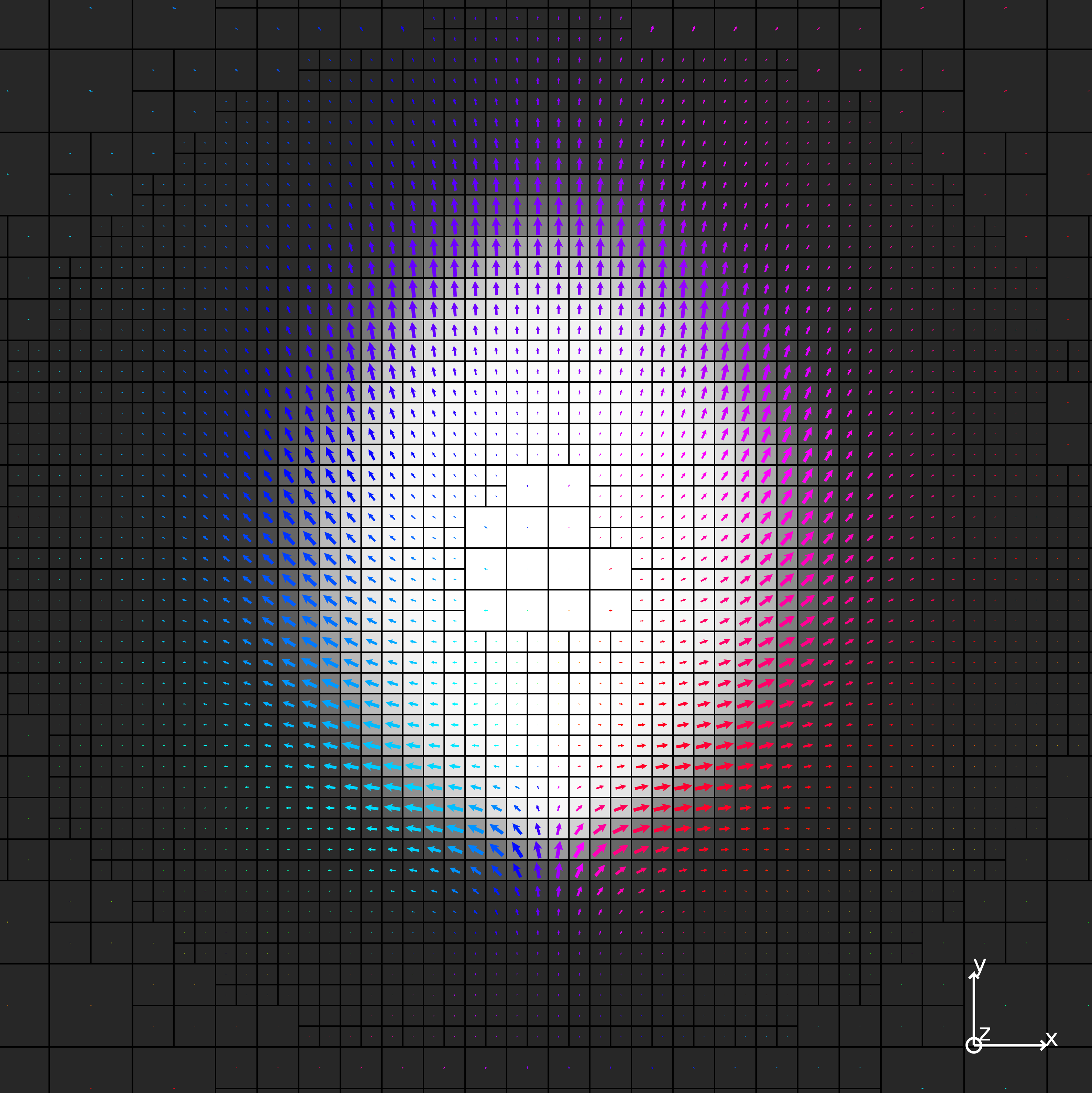}
  
  \vspace{0.2cm}
  
  \includegraphics[width=0.48\linewidth]{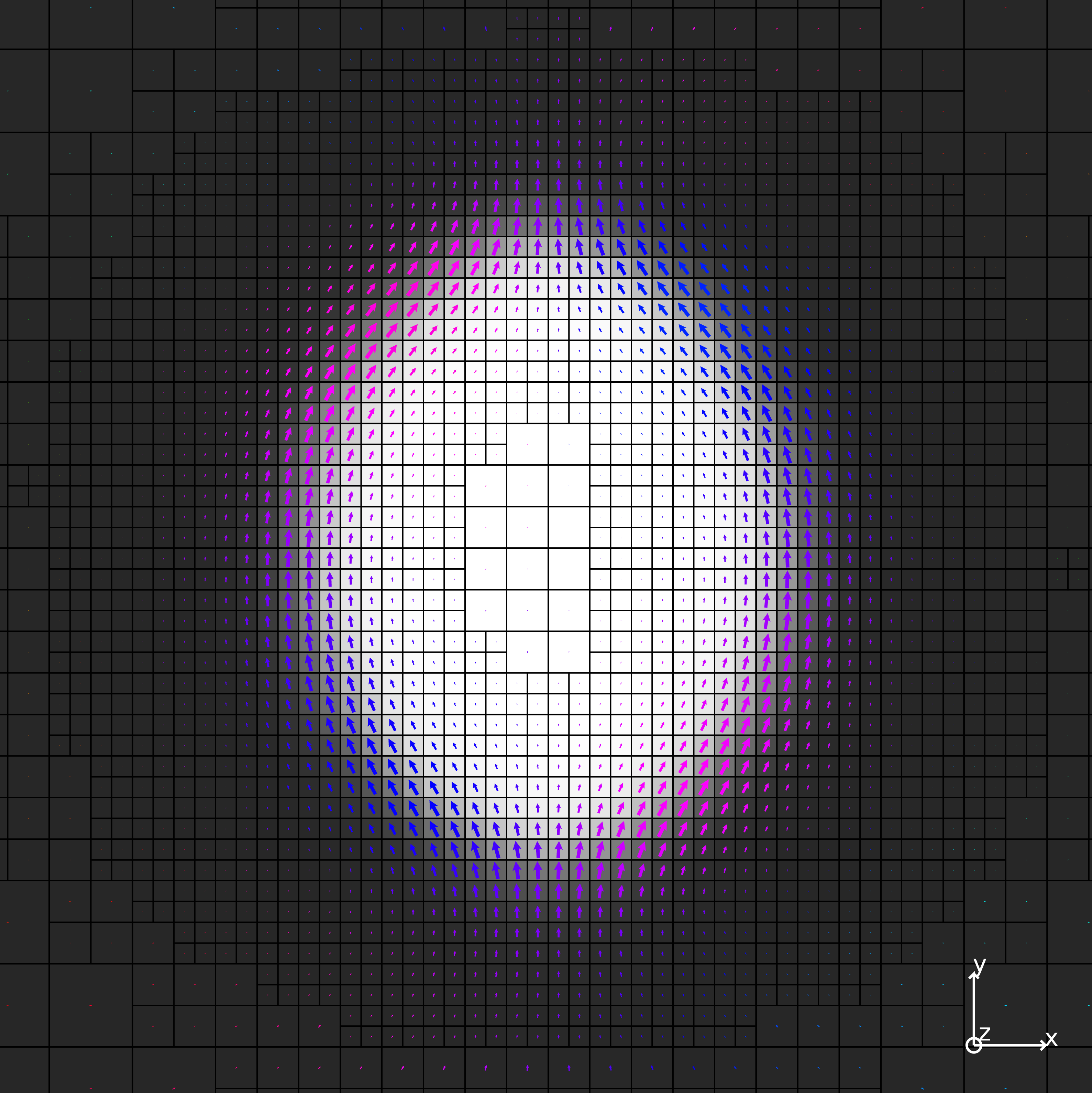}
  \hspace{\fill}
  \includegraphics[width=0.48\linewidth]{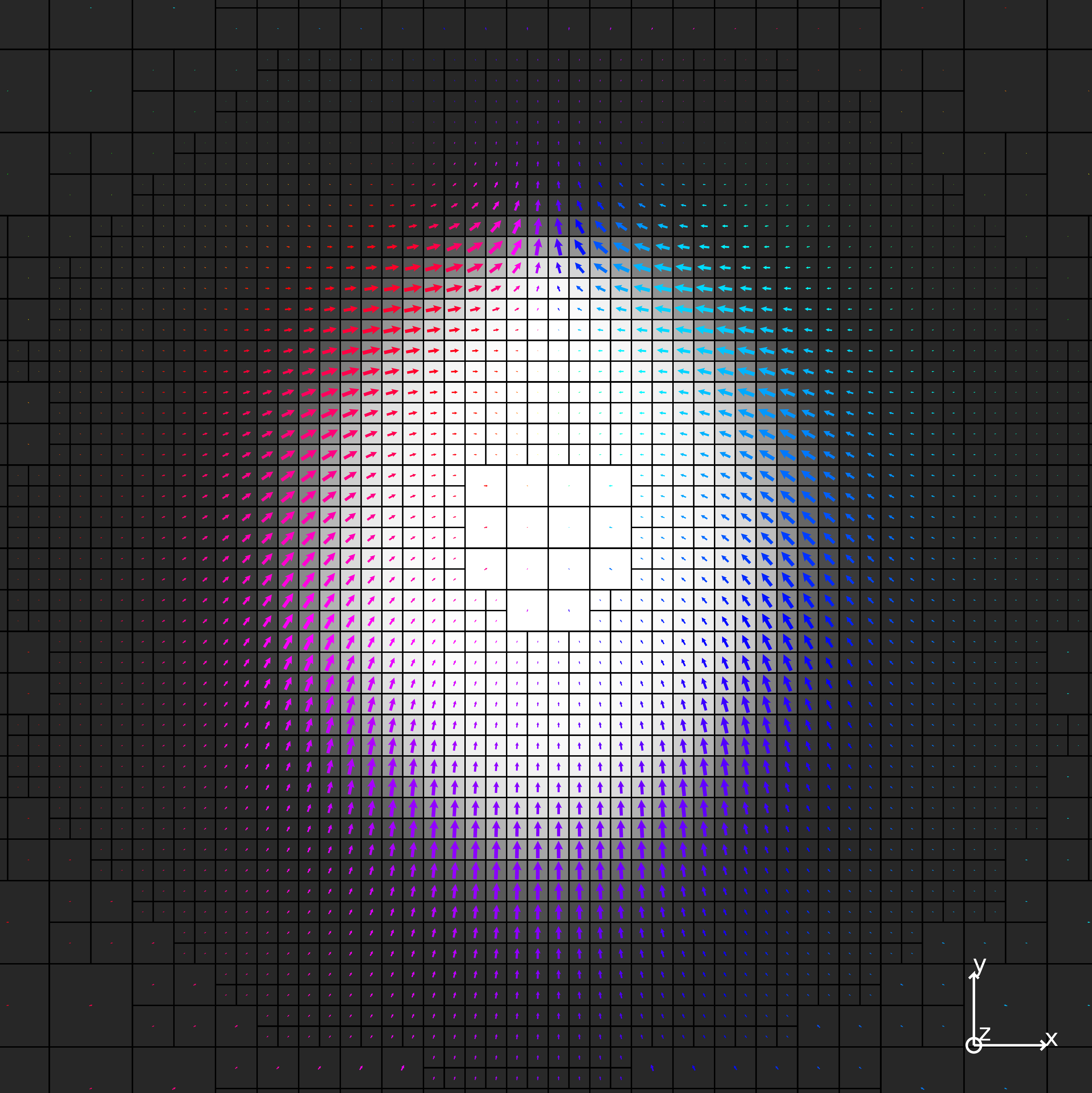}  

  \caption{(Color online) Cross-sections of a system containing two
    VBL with no BP. From top left to bottom right: whole system at
    $z=h/2$ (lateral size $218~\textrm{nm}\times 218~\textrm{nm}$),
    zoom at $z=h$, $z=h/2$ and $z=0$ (lateral size
    $90~\textrm{nm}\times 90~\textrm{nm}$). The system is 15~nm thick
    and a field of 0.25~T is applied.}
  \label{fig-bubble-without-BP}
\end{figure}

For a thickness of 15~nm, the configuration containing VBL with BP is
not stable and the two BP migrate towards the two opposite surfaces of
the system. The two regions that exbibit high spatial variations of
the magnetization (360$^\circ$ rotation for straight domain walls) are
thus located on opposite sides of the system
(Fig.~\ref{fig-bubble-without-BP}). This disappearance of the two BP
is associated with a deformation of the domain wall, in agreement with
the one found on straight domain walls in the previous section and
with experimental results. Likewise the charges are minimized and the
exchange energy decreases.

It is worth noting that the magnetization in the two lines is oriented
in the same direction. This is called the winding
configuration\cite{hubert98}.  Lines with opposite orientations of the
magnetization constitute the unwinding configuration, and have found
to be unstable: the two lines annihilate and the bubble is
circular. Indeed, in order to minimize charges in both VBL the bubble
would have a ``heart''-like shape, which is not favorable. The
orientation in the two lines is close to the orientation in the rest
of the domain wall at $z=h/2$.


\begin{figure}[htbp]
  \centering
  \includegraphics[width=\linewidth]{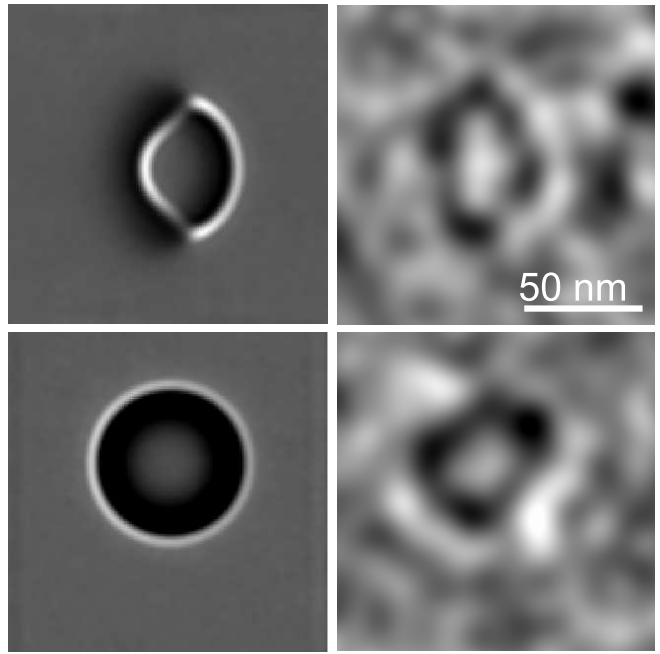}
  \caption{Comparison of simulated Fresnel contrasts and experimental contrasts for the two type of bubbles observed. The defocalisation used is 100~$\mu$m.}
  \label{LTEM2}
\end{figure}

A further step can be made towards the comparison between simulated
and experimental configurations by simulating Fresnel contrasts that
would be obtained from the multiscale calculations. They are given in
Fig.~\ref{LTEM2}. Beside the result corresponding to
Fig.~\ref{fig-bubble-without-BP}, we report the results for a
bubble without a BP. It can be seen that the position of the contrast
and the shape of the bubble agree fairly well.



Despite the good agreement on the shape of bubbles, the transition
from the BP-free to the BP configuration does not occur at the same
thickness experimentally and in the simulations. Indeed, the
configurations without BP are not stable in our simulations for a
thickness of 37.6~nm (and even 20.7~nm), whereas according to the
deformation of the bubbles observed in the samples, VBL contain a BP
at this thickness. One reason for this discrepancy may be the presence
of the soft layer on which the L1$_0$ layer is deposited. The exchange
and demagnetizing contributions to the energy are modified due to the
different closure of the magnetic flux. The thickness of the bottom
N\'eel cap increases\cite{Masseboeuf2008}, which induces a dissymmetry in
the system and could favor the configuration without BP.



\section{Conclusion}
\label{sec:conclusion}

Using Lorentz transmission electron microscopy on FePd samples and
multiscale simulations, we have shown that it is possible to determine
the magnetic structure of domain walls as thin as 8~nm. The presence
of vertical Bloch lines in some bubbles has been demonstrated by
microscopy. Bubbles containing two vertical Bloch lines exhibit a
distortion of the classical circular shape. The simulation of entire
bubbles has been possible thanks to the multiscale approach and has
revealed that the deformation observed experimentally is a signature
of the absence of Bloch points inside the vertical Bloch lines. For
straight domain walls in FePd, we predict a larger buckling than
previously reported for other materials.


\bibliography{Magnetisme,CG_Magnetisme,MultiScale,VBL,LTEM}

\end{document}